\documentclass[twocolumn,showpacs,preprintnumbers,amsmath,amssymb]{revtex4}

\usepackage{graphicx}
\usepackage{dcolumn}
\usepackage{bm}

\begin{document}

\preprint{APS/123-QED}

\title{Coherent resonant interactions and slow light with molecules confined in photonic band-gap fibers}

\author{Saikat Ghosh}
\author{Jay E. Sharping}%
\author{Dimitre G. Ouzounov}
\author{Alexander L. Gaeta}%
 \email{a.gaeta@cornell.edu}
\affiliation{%
School of Applied and Engineering Physics\\
Cornell University\\
Ithaca,NY 14853}%

\date{\today}

\begin{abstract}
We investigate resonant nonlinear optical interactions and
demonstrate induced transparency in acetylene molecules in a
hollow-core photonic band-gap fiber at 1.5$\mu$m. The induced
spectral transmission window is used to demonstrate slow-light
effects, and we show that the observed broadening of the spectral
features is due to collisions of the molecules with the inner
walls of the fiber core.  Our results illustrate that such fibers
can be used to facilitate strong coherent light-matter
interactions even when the optical response of the individual
molecules is weak.
\end{abstract}

\maketitle


\par The fields of photonics and quantum optics have undergone immense
technological and scientific advancement in the past decade. The
growth of the telecommunications industry has given rise to exotic
devices such as the hollow-core photonic band-gap fiber
(HC-PBF)~\cite{russell1,HCPBG1.5micron} which can guide light over
hundreds of meters through a hollow core surrounded by a photonic
crystal structure (see Fig.1.) that localizes light in the core.
In addition, since the nonlinearities are determined primarily by
the gas in the fiber's core, such fibers can be used to produce
efficient Raman scattering~\cite{russell2} and solitons with
megawatt powers~\cite{Ouzounov}.

\par In parallel with these developments, there has been tremendous progress in the
field of coherent atom-photon interactions. The richness of
physics in manipulating quantum states of matter and creating
coherent superpositions has led to numerous studies in atom
cooling and Bose-Einstein condensation~\cite{BEC}, quantum
information and computation~\cite{Quantum information1,Quantum
information2,Quantum information3}, and electromagnetically
induced transparency (EIT)~\cite{EIT1, EIT2}. In particular, the
phenomenon of EIT has led to an assortment of potentially
practical applications such as ultra-slow light~\cite{slow light1,
slow light2}, light storage in a medium~\cite{stored light2}, and
single-photon switching~\cite{schmidt96,single photon switching}.
Our research is motivated by the desire to integrate novel
photonic devices~\cite{schmidt2} with coherent atom-photon
interactions and to build useful devices for the fields of
telecommunications and quantum information processing. Hollow-core
photonic band-gap fiber, with the ability to confine atoms and
molecules to its core for extremely long interaction lengths and
to integrate itself readily with the existing telecommunication
technology, seems to be the ideal bridge between photonics and
quantum optics.

\par   In this Letter, we
describe experiments that represent the first demonstration of
coherent resonant interactions in these fibers. We perform
coherent three-level spectroscopy of acetylene molecules confined
to the core of a HC-PBF. Using a strong control beam to modify
coherently a three-level molecular system, we observe induced
transparencies of the probe beam as large as 50\% and we
demonstrate slow light delays via this transparency window. These
results indicates that such fibers can be used to greatly enhance
light-matter interactions even when the optical response of the
individual molecules is weak.

\par Acetylene has attracted significant
interest due to its spectral overlap with the low-loss C-band
fiber telecommunications window~\cite{gilbert}. In addition,
several low-loss HC-PBF have been demonstrated with bandgaps in
this spectral range~\cite{HCPBG1.5micron}. This motivates us to
study coherence effects in acetylene ($C_2H_2$), a linear
symmetric molecule, which has clean ro-vibrational
transitions~\cite{c2h2} in this wavelength range~\cite{ritari04}.
Only a few studies have demonstrated coherence effects in
molecular systems~\cite{molecules-coherence1,
molecules-coherence2}, and one of the primary reasons is due to
the typically weak oscillator strengths of the molecular
transitions. Observation of an appreciable effect in a bulk
molecular gas necessitates the opposing requirement of a tightly
focused laser beam and a long interaction length. While the
effective interaction between the optical field and the molecular
system can be enhanced by using an optical cavity, the small core
diameter of the HC-PBG, together with practically loss-less
propagation of fields over many meters, provides an attractive
alternative.

\begin{figure}
\includegraphics[width=3in]{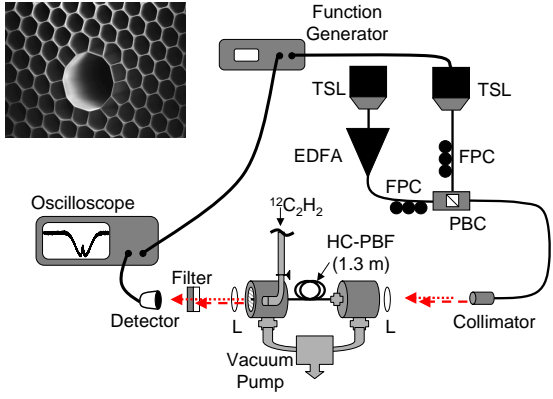}
\caption{ Schematic diagram of the experimental setup. The
transverse cross-section of the HC-PBF is also shown. EDFA:
erbium-doped fiber amplifier; TSL: tunable semiconductor laser;
FPC: fiber-polarization controller; PBC: polarization-beam
combiner; L: lens.} \label{fSetup}
\end{figure}

\par In our experiments we use a 1.33-m-long HC-PBF (Blazephotonics, HC-1550-01) which
has a core diameter of 12\,$\mu$m and a band gap extending from
1490 to 1620\,nm. As shown in Fig.~\ref{fSetup}, the two ends of
the fiber are placed in separate vacuum cells. During assembly, we
purge the air inside the fiber using nitrogen gas and then
evacuate to a pressure below $10^{-6}$ Torr. One of the cells is
then filled with 99.8\% pure $C_2H_2$. We detect the presence of
the gas in the other cell spectroscopically to ensure that the
core is filled with acetylene. All experiments are performed in
steady state with both of the cells at the same pressure, which
ranges from 10-100\,mTorr. For the fiber used in these
measurements, steady state is reached in less than an hour.

\par We use two tunable external-cavity lasers for pump-probe spectroscopy. The emission from one
of the lasers is tuned to 1535\,nm and amplified by a 1-W
erbium-doped fiber amplifier (EDFA). We use the amplified beam as
our control beam. The other laser is tuned to 1517\,nm and serves
as the probe beam. The control and probe beams are combined using
fiber polarization controllers and a polarization beam combiner.
The co-propagating control and probe beams are collimated and then
launched into one end of the fiber with coupling efficiencies of
50\% using a microscope objective. We filter out the control beam
from the probe using an interference filter with a rejection ratio
of $>$50\,dB, and the probe is detected by a photodetector.

\par
\begin{figure}
\includegraphics[width=3in]{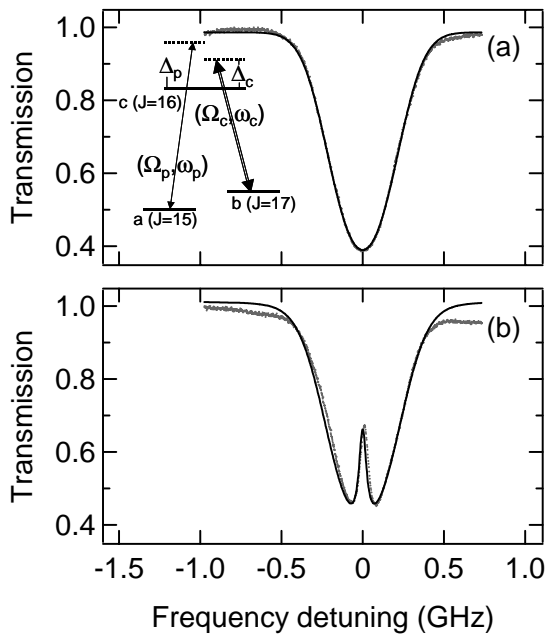}
\caption{(a) Measured (points) and theoretical (line)
Doppler-broadened absorption spectra for the R(15) transition in
acetylene within a 1.3-m segment of photonic-band-gap fiber. A
diagram of the relevant molecular levels are shown in the inset.
(b) Measured (points) and theoretical (line) spectra for the same
transition in the presence of a strong control beam (320\,mW at
the fiber output).} \label{absorptions}
\end{figure}

 The energy-level scheme for the molecules used in our experiment is shown in Fig.~\ref{absorptions}.
The control beam is tuned between the levels
0(${\Sigma^{+}}_g$)(J=17) and
$\nu_1$+$\nu_3$(${\Sigma^{+}}_u$)(J=16), denoted as levels $b$ and
$c$, respectively. The probe beam is tuned between level $a$,
which is 0(${\Sigma^{+}}_g$)(J=15), and $c$.  These are the P(17)
and R(15) lines of $^{12}C_2H_2$ at 1535.3927 nm and 1517.3144 nm
respectively~\cite{c2h2}. The probe power is maintained below
500\,$\mu$W, and the wavelength of the probe is scanned over the
R(15) transition line. As shown in Fig.~\ref{absorptions}(a),
without the control beam we observe the Doppler-broadened
absorption of the probe beam with a width of 480\,MHz. In the
presence of the control beam, a transparency window is opened.
Figure~\ref{absorptions}(b) shows a typical trace of the
probe-field absorption of the R(15) line in presence of a 320-\,mW
(measured at the output of the fiber) control beam tuned exactly
to the center of the P(17) transition. While induced
transparencies of this magnitude have been observed routinely in
focused geometries in atomic systems such as rubidium, observing
this transparency feature in acetylene is remarkable in light of
the fact that the oscillator strength of the transition is several
orders of magnitude less than that for the $D$ lines of rubidium
transitions.

\par For an understanding of our experimental results, we solve the
density matrix equations for a three-level $\Lambda$ system (see
Fig. \ref{absorptions}), where one of the two lower levels,
$(J_b,m)$ is coupled to the upper level $(J_c,m)$ with a strong
control field $E_c$, while the transition $(J_a,m)-(J_c,m)$ is
probed with a weak field $E_p$. Here $m$ is the orientational
quantum number, denoting the alignment of the otherwise degenerate
ro-vibrational levels. We first calculate the steady-state
population distribution for the three-level system in presence of
the strong control field and no probe field~\cite{scully}. We
consider a semi-open system, where the upper level radiatively
decays to the two lower levels with a decay rate $(\gamma)$
independent of $m$. We further consider a collisional
redistribution of population between the two lower levels at a
rate $w_{ab}=w_{ba}$. We use the resulting steady-state population
expressions $\rho^{m(0)}_{ii}$ for levels $i = a, b, c$, to
calculate the coherence $\rho^m_{ca}$ correct to first order in
the probe field,
\begin{widetext}
\begin{equation}
\rho^m_{ca}=\frac{-i\Omega^m_p}{2(\gamma_{ca}-i\Delta_p+\frac{|\Omega^m_c|^2/4}{\gamma_{ba}-i(\Delta_p-\Delta_c)})}[(\rho^{m(0)}_{cc}-\rho^{m(0)}_{aa})+\frac{|\Omega^m_c|^2}{4(\gamma_{cb}+i\Delta_c)[\gamma_{ba}-i(\Delta_p-\Delta_c)]}(\rho^{m(0)}_{bb}-\rho^{m(0)}_{cc})],
\label{rhom}
\end{equation}
\end{widetext}
where $\Delta_c=\omega_c-\omega_{cb}-k_cv$ and
$\Delta_p=\omega_p-\omega_{ca}-k_pv$ are the velocity-dependent
detunings of the control and the probe field, with wave vectors,
 $k_c$ and $k_p$, respectively, and $\Omega^m_c$ is the Rabi frequency of the control field,
which can be expressed as
\begin{eqnarray}
\Omega^m_c&&=\frac{2\langle\Sigma^+_u(J_c,m)\mid\hat{\mu}\mid\Sigma^+_g(J_b,m)\rangle
E_c}{\hbar}\nonumber\\
&&=\frac{2\mu^0F^0_{\Sigma-\Sigma}(J_c,m;J_b,m)E_c}{\hbar},
\label{omegam}
\end{eqnarray}
where $\mu^0$ is the pure electronic-vibrational transition dipole
moment, and $F^0_{\Sigma-\Sigma}(J_c,m;J_a,m)$ is the
H\"{o}nl-London factor for the transition~\cite{spano}. Similarly,
$\Omega^m_p$ is the Rabi frequency associated with the probe
field. The control and probe beams are assumed to be co-polarized.
Equation~(\ref{rhom}) reduces to the usual EIT coherence for a
closed three-level system, with the approximation of
$\rho^{m(0)}_{aa}\simeq1,\rho^{m(0)}_{bb}=\rho^{m(0)}_{cc}\simeq
0$. The first term inside the parentheses contributes to
saturation of the transition $(J_b,m)-(J_c,m)$ due to the presence
of the control field with powers above the saturation
intensity~\cite{javan}, and the second term represents an
interference due to the oscillation of population between
$(J_b,m)-(J_c,m)$.

The dephasing rates are expressed in the form,
$\gamma_{ij}=(\gamma_i+\gamma_j)/2+\gamma^{coll}_{ij}$, for
$i=a,b,c$, where, $\gamma_i$ is the decay rate of level $i$ and
$\gamma^{coll}_{ij}$ is the dephasing rate due to collision of the
molecules with each other and with the inside wall of the HC-PBG
fiber. Following Ref.~\cite{nakagawa}, an estimate for the
pressure-broadened lifetimes for these transitions of $C_2H_2$ at
wavelengths of 1.5\,$\mu$m is of the order of 10\,MHz/Torr. At our
working pressures near 50\,mTorr, the line width is estimated to
be less than 1\,MHz. In contrast, an estimate of the linewidth due
to dephasing collisions with the inside walls of the fiber,
assuming a mean thermal velocity of the molecules of 400\,m/sec
inside a cylinder of diameter 12\,$\mu$m, predicts a value which
is an order of magnitude larger. We expect these collisions to be
the major source of decoherence in our system. The dephasing rate
$\gamma_{ij}$ is therefore assumed to be the same for all
coherences, with $\gamma^{coll}_{ij}=\gamma^{coll}_{wall}$.

\par To compare our experiment with theory, we first average the
coherence over the Doppler profile and sum over the orientation
({\it m}) to obtain the probe field susceptibility $\chi$, and we
fit our experimental data to the transmission coefficient
\begin{equation}
T=\frac{I_{out}}{I_{in}}=exp[-4\pi k_pL ~\textrm{Im}\{\chi\}],
\label{tcoeff}
\end{equation}
where $L$ is the length of the fiber. In Eq.~(\ref{tcoeff}), we
assume the Rabi frequency for the control field is constant along
the length of the fiber.
\par The variable parameters in the theory are: (i) the constant term in the exponent of Eq.~(\ref{tcoeff}), which
is the product of the number density, the length of the fiber, and
the constants for the Rabi frequency of the probe field, (ii) the
collisional redistribution rate of the ground-state population
$w_{ab}$, (iii) the pure electronic-vibrational Rabi strength
$\Omega^0_c=2\mu^0E_c/ \hbar$, (iv) the upper-state decay rate
$2\gamma$, and (v) the dephasing rate $\gamma^{wall}_{coll}$ due
to collisions with the inside walls of the fiber. We first
estimate the constant for (i) for the probe field by fitting the
Doppler-broadened probe-field absorption without the control field
(Fig. 2a.). The relative height of the probe-field absorption,
with and without the control field, depends primarily on the rate
$w_{ab}$. A choice of $w_{ab}=0.4\gamma$ matches well with all our
experimental absorption curves over the entire range of the
control-field strength. We fit the remaining parameters,
$\Omega^0_c$, $\gamma$, and $\gamma^{wall}_{coll}$ to the
experimental data. We make an initial guess of $\Omega^0_c$ by
estimating $E_c$ from the input control field power in the fiber
and using the values for $\mu^0$ from Ref.~\cite{auwera}. For
$\gamma$ we use an estimate from Ref.~\cite{nakagawa}, and our
initial guess for $\gamma^{wall}_{coll}$ is as discussed above.
\par
\begin{figure}
\includegraphics[width=3in]{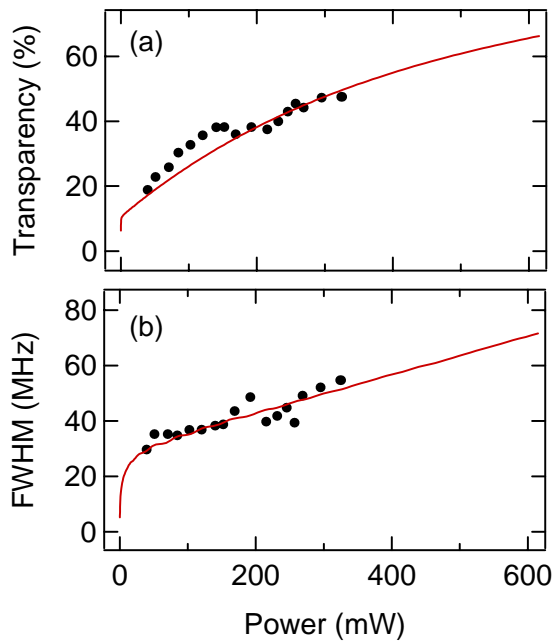}
\caption{(a) Experimental (circles) and theoretical (lines)
relative transparencies (ratio of the change in the zero-detuning
probe transmission in the presence of the control beam to the
zero-detuning probe transmission without the control beam) as
functions of control power measured at the output of the fiber.
(b) The corresponding FWHM of the transparency dip.}
\label{transparencies}
\end{figure}

 As seen in Fig. ~\ref{absorptions}(b), our theoretical model is in excellent
agreement with our experimentally measured probe-field absorption.
Our value for $\Omega^0_c=9.3$\,MHz, is consistent with the
measured control power of 320\,mW, and a value of
$\gamma=0.96$\,MHz and $\gamma^{coll}_{wall}=17.28$\,MHz is
estimated from the fit. Figure~\ref{transparencies}(a) shows a
plot of the transparency as a function of the control power
measured at the output end of the fiber together with the
corresponding theoretical plot. Presently in our experiments, we
are limited by the power of our amplifiers for the control field,
but theory predicts a much larger transparency at higher powers.
The measured transparency full-width at half maximum (FWHM) in
Fig.~\ref{transparencies}(b) shows a linear dependence on control
power, which has the behavior of the EIT linewidth
$\Gamma_{EIT}\Rightarrow {\Omega^0_c}^2$ of Ref.~\cite{scully}.

\par
\begin{figure}
\includegraphics[width=3in]{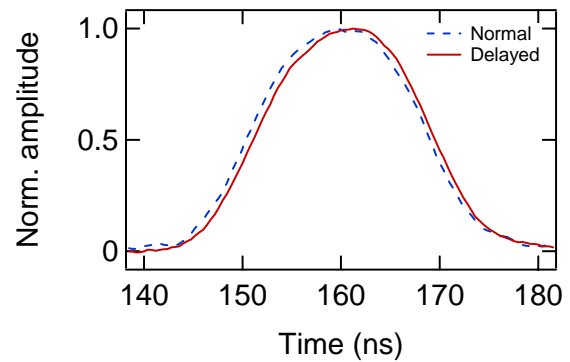}
\caption{Normalized amplitude vs. time for a probe pulse with
(delayed) and without (normal) the control beam. See text for
measurement details.} \label{delay}
\end{figure}

 An interesting application of our system, which has relevance to both telecommunications and quantum
information, is to use this spectroscopic feature to induce a
controllable pulse delay. Sharp changes in the absorption of a
probe field due to coherent effects produced by a  strong control
field are accompanied by dramatic changes in the refractive
index~\cite{slow light1, slow light2}. The group delay experienced
by a probe pulse, can be expressed in the form
\begin{equation}
T_D=\frac{L}{v_g}-\frac{L}{c}=2\pi\omega_p\frac{L}{c}\left[
\frac{\delta
~\textrm{Re}\{\chi\}}{\delta\omega_p}\right]_{\omega_p=\omega_{ca}}.
\end{equation}
One of the major advantages of our system is that, for a given
$L$, we can produce large delays even at low pressures. As a proof
of concept, we generate 19-ns (FWHM) probe pulses using an
amplitude modulator and tune the wavelength of the probe to be on
resonance with the $a-c$ transition. The wavelength of the
copropagating CW control beam is tuned to be on resonance with the
$b-c$ transition. The control beam is filtered using a tunable
bandpass filter, and the probe pulses are detected with a 10-\,GHz
receiver. For these measurements the control beam power measured
at the output of the fiber is 200\,mW, and the on-resonance probe
absorption and relative transparency are both approximately 40\%.
Figure~\ref{delay} shows the probe pulse with the control beam off
and with it on. The delay measured in the presence of the control
beam is 800\,ps. This represents, to our knowledge, the first
demonstration of slow light at telecommunications wavelengths
using molecular spectroscopic features.

\par In summary, we have investigated coherent 3-level resonant interactions with acetylene
molecules inside a HC-PBF. A
theoretical model was used to estimate the effect of decoherence
due to collision of the molecules with the inside wall of the
fiber. We used this spectroscopic feature to demonstrate a
potential application of producing slow light with molecules
inside optical fibers at telecom wavelengths.
\par We thank K. Moll, K. Koch and D. Gauthier for useful discussions and comments. We
gratefully acknowledge support by the Center for Nanoscale
Systems, supported by NSF under award number EEC-0117770, the Air
Force Office of Scientific Research under contract number
F49620-03-0223, and DARPA under the Slow Light program.


{}
\end{document}